\documentclass[12pt,preprint,eqsecnum]{article}
\usepackage{color}
\usepackage{amsmath, amssymb}
\usepackage{epsfig, concmath, palatino}
\usepackage{pstricks,pst-node,pst-tree}
\usepackage{cite}

\newcommand{\Mellin}[1]{{\cal M}\left[#1\right]}
\newcommand{\invMellin}[1]{{\cal M}^{-1}\left[\,#1\,\right]}

\def\cS{{\cal S}}

\def\tcP{\tilde{\cal P}}
\def\half{{\textstyle {\frac12}}}
\def\hA{\widehat{\cal A}}
\def\hB{\hat{\cal B}}
\def\cP{{\cal{P}}}
\def\gone{\gamma}
\def\dgone{\dot{\gone}}
\def\ddgone{\ddot{\gone}}
\def\dddgone{\dddot{\gone}}
\def\dcP{\dot{\cP}}
\def\ddcP{\ddot{\cP}}
\def\dddcP{\dddot{\cP}}

\def\cO#1{{\cal O}\left(#1\right)}
\def\ph{{\mbox{\scriptsize ph}}}
\def\alp{\mathbf{\alpha}_{\mbox{\scriptsize ph}}}
\def\gph{\mathbf{g}_{\mbox{\scriptsize ph}}}

\def\hcS#1{\widehat{\cal S}_{#1}}
\def\ucS#1{\underline{{\cal S}_{#1}}}
\def\tucS#1{\underline{\tilde{\cal S}_{#1}}}
\def\hucS#1{\underline{\widehat{\cal S}_{#1}}}
\def\tcS#1{\tilde{\cal S}_{#1}}

\def\vec#1{\mathbf #1}
\def\hPhi#1{\widehat{\Phi}_{#1}}
\def\tPhi#1{\tilde{\Phi}_{#1}}

\title{Twist 3 of the $sl(2)$ sector of $N=4$ SYM\\
and reciprocity respecting evolution} 
\author{M.\ Beccaria$^1$, Yu.L.\ Dokshitzer$^2$\footnote{On leave of absence:
 St.\ Petersburg Nuclear Physics Institute, 188350, Gatchina, Russia}  
    { } and G.~Marchesini$^3$   \\
  \normalsize
  $^1$ Physics Department, Lecce University and INFN Sezione di Lecce, Lecce, Italy \\
   \normalsize  $^2$LPTHE, Universities of Paris--VI--VII and CNRS, Paris, France\\
  \normalsize $^3$University of Milano--Bicocca and INFN Sezione di
  Milano--Bicocca, Milan, Italy}
\date{}

\begin{document}

\maketitle

\vspace{-8cm}
\begin{flushright}
\end{flushright}
\vspace{6cm}

\abstract{We consider the bosonic $sl(2)$ sector of the maximally supersymmetric ${\cal{N}}\!=\!4$ SYM model and show that anomalous dimension 
of the twist-3 single-trace composite operators built 
of scalar fields, recently calculated  up to the four-loop order, 
can be generated by a compact reciprocity respecting evolution kernel.


\section{Introduction}

QCD shares the vector boson --- gluon --- sector with supersymmetric Yang--Mills models (SYM).
This suggests to explore supersymmetric partners of QCD in order to shed light 
on the subtle structure of the perturbative quark--gluon dynamics near the light cone. 
The latter manifests itself in parton distributions (structure functions 
of deep inelastic lepton--hadron scattering, DIS) and parton$\to$hadron fragmentation functions  
(internal structure of jets in, e.g., $e^+e^-$ annihilation into hadrons)
which evolve with the hardness scale of the process, $Q^2$. 

QCD is not an integrable quantum field theory. 
In spite of this, in certain sectors of the chromodynamics 
the integrability does emerge \cite{Lip93}. This happens, markedly, in the problem of 
high energy Regge behaviour of scattering amplitudes in the large $N_c$ 
approximation (planar `t Hooft limit),  
in the spin $\tfrac32$ baryon wave function, 
for the scale dependence of specific (maximal helicity) quasi-partonic  
operators (for review see \cite{Belitsky:2004cz}). 
What all these problems have in common, is the irrelevance of quark degrees 
of freedom and the dominance of the {\em classical}\/ part of gluon dynamics, 
in the sense of the Low--Burnett--Kroll theorem~\cite{LBK}.

A SYM dynamics exhibits the deeper integrability, the higher the symmetry. 
The maximally supersymmetric $\mathcal{N}\!=\!4$ YM theory 
occupies an exceptional position in this respect. 
A string of recent impressive theoretical developments 
\cite{Belitsky:2004cz,Beisert:2005fw,impressive}
hints at an intriguing possibility that this QFT, 
super-conformally invariant at the quantum level, may admit 
an all loop solution for anomalous dimensions of its composite operators. 
QCD would benefit a lot from such a solution, since this would provide 
a one-line-all-loops 
representation for the {\em dominant}\/ part of the perturbative gluon dynamics, 
the part that could be dubbed {\em LBK-classical}. 

For the non-compact $\mathfrak{sl}(2)$ sector
containing twist operators with arbitrarily high spin the three-loop Bethe Ansatz 
equations (BAE) have been conjectured in~\cite{Staudacher:2004tk}.  
The perturbative expansion of the Staudacher BAE 
perfectly matches the {\em three-loop}\/ anomalous dimensions 
of twist-2 Wilson operators that determine the Bjorken scaling violation pattern. 
At the same time, it apparently fails at the {\em fourth loop}\/ where 
the wrapping problem is manifest \cite{KLRSV:2007}.

\medskip

Since in $\mathcal{N}\!=\!4$ SYM all twist-2 operators belong to one super-multiplet, 
diagonalization of the matrices 
$\gamma (3\times 3)$ and $\tilde\gamma(2\times2)$ describing evolution of unpolarized ($g,\lambda,\phi$)
and polarized parton distributions ($g,\lambda$) results in the relation \cite{Lipatov:2000}
\begin{equation}\label{eq:chain}
\gamma_+(N+2) = \tilde\gamma_+(N +1) = \gamma_0(N) = \tilde\gamma_-(N- 1) = \gamma_-(N- 2) 
\>\equiv\> \gamma_{\mbox{\scriptsize uni}}(N).
\end{equation}
It expresses all five diagonal elements 
in terms of the unique function --- the ``universal anomalous dimension'' 
$ \gamma_{\mbox{\scriptsize uni}}$ with shifted arguments.  
Inspired by the structure of the answer in the first two loops 
\cite{Dolan:2000ut,Arutyunov:2001mh,KotLip:2001,KotLipVel:2003}, 
Kotikov, Lipatov, Onishchenko and Velizhanin (KLOV) have proposed 
in~\cite{Kotikov:2004er} 
the {\em maximum transcendentality principle}\/ according to which 
$\gamma^{(n)}_{\mbox{\scriptsize uni}}$ at $n$ loops is a linear combination 
{\em Euler--Zagier harmonic sums}\/ of transcedentality $\tau=2n-1$. 
This allowed them  to predict $\gamma^{(3)}_{\mbox{\scriptsize uni}}$ 
by simply picking up from the three-loop non-singlet QCD anomalous dimension 
\cite{Moch:2004pa} the ``most transcendental'' terms ($\tau= 5$), having set $C_F=C_A=N_c$. 

\medskip
The complexity of higher loop expressions for $\gamma^{(n)}_{\mbox{\scriptsize uni}}$, $n\ge 2$, 
is significantly reduced if one trades the anomalous dimension for a new object 
--- the reciprocity respecting (RR) evolution kernel, \cite{Dokshitzer:2006nm}. 
In the case of $\mathcal{N}\!=\!4$ SYM ($\beta(\alpha)\equiv 0$) 
this trade-off takes an exceptionally simple form,  
\begin{equation}\label{eq:gamma-P}
   \gamma_\sigma(N) = \cP(N - \half \sigma\gamma_\sigma(N)), \quad 
   \sigma = \left\{ \begin{array}{ll} +1, & \mbox{T} \\
    -1, & \mbox{S} \end{array}\right. ,
\end{equation}
where $\gamma_\pm(N)$ are the anomalous dimensions responsible for 
time- and space-like parton evolution, respectively 
(in the S case which admits the operator product expansion, 
 $\gamma_-$ describes the scale dependence of Wilson operators of the leading twist 2).  
Here we employ the definition of the anomalous dimension which is more familiar to the integrable 
community\footnote{The r.h.s.\ of \eqref{eq:violation} is traditionally called 
$\gamma$ in the QCD parton evolution context.}; 
within this convention the scaling violation rate is given by the expression 
\begin{equation}\label{eq:violation}
    \frac{d}{d\ln Q^2} \ln D_N(Q^2) = -2\,\gamma(N)\, ,
\end{equation}
with $D_N(Q^2)$ the Mellin moment of the parton distribution measured 
at the large momentum transfer (hardness) scale $Q^2$.

In the context of the QCD parton picture, 
the notion of the RR evolution kernel $\cP$ emerges as a result of the reformulation 
of space-like (DIS, ``S'') 
and time-like parton multiplication processes ($e^+e^-$, ``T'')
in terms of a unified evolution equation~\cite{Dokshitzer:1995ev,Dokshitzer:2005bf}. 
This equation is constructed on a basis of  the {\em parton fluctuation time}\/ ordering. 
In the space of Bjorken (S)/Feynman (T) variable $x$ conjugate to the Lorentz spin $N$,
\begin{equation}
 \cP(N) = \int_0^1\frac{dx}{x}\,x^N\cP(x) \equiv \Mellin{\tcP(x)},
\end{equation}
 its integral kernel $\tcP(x)$,  identical for the two channels, satisfies 
the Gribov--Lipatov reciprocity (GLR) \cite{Gribov:1972} in all orders:
\begin{equation}\label{eq:GLR}
   F(x) \>=\>  -x\,F(x^{-1}) . 
\end{equation}
Existence of the common evolution kernel $\cP$ and its
internal symmetry is reflected in the structure of the anomalous dimensions 
$\gamma_\sigma$ that one finds solving \eqref{eq:gamma-P}. 

Basso and Korchemsky in  \cite{Basso:2006nk} approached this problem from 
the point of view of the large $N$ expansion.  
They have generalised \eqref{eq:gamma-P} to anomalous dimensions 
of quasi-partonic operators of arbitrary twist $L$, and
traced its origin to the conformal symmetry.

Having analysed all anomalous dimensions of twist-2 operators, known to the two- 
and/or three-loop order in QCD and SYM theories, as well as in the scalar $\lambda\phi^4$ QFT (known at four loops), the authors demonstrated that the asymptotic series for the corresponding
kernels $\cP(N)$ run in integer negative powers of $N(N\!+\!1)$ 
--- the quadratic Casimir of the collinear ${SL}(2;\mathbb{R})$ group.
As a result, even, $N^{-2n}$,  and odd terms, $N^{-2n-1}$, 
of the large $N$ expansion of {\em anomalous dimensions}\/ $\gamma_\sigma$ 
turn out to be related.  
Basso and Korchemsky named this feature ``parity preserving asymptotic series''. 
 
In $\mathcal{N}\!=\!4$ SYM the property of parity preserving series directly follows 
from the representation of the twist-2 kernel
developed in three loops in \cite{Dokshitzer:2006nm}: 
\begin{equation}\label{eq:S1only1}
 \cP(N) = 4\,S_1(N)\cdot \bigg(\frac{N_c\alp}{2\pi} + \hA(N) \bigg) + {\cal B}(N).
\end{equation} 
Here $\alp$ is the physical coupling which determines 
the strength of LBK-classical radiation \cite{CMW,Dokshitzer:1995ev} 
and coincides with the so-called cusp anomalous
dimension \cite{Korchemsky:cusp} 
whose all-order weak-coupling expansion  in $\mathcal{N}\!=\!4$ SYM 
has been remarkably guessed in~\cite{Beisert:2006ez}. 

The functions $\hA=\cO{\alpha^3}$ and $\hB=\cO{\alpha^2}$ 
have compact expressions in terms of RR combinations of 
complementary nested harmonic sums. Their asymptotic expansion at $N\to +\infty$
is {\em regular}, that is, contains no $\ln^p N$ enhancement factors 
at any level of the $1/N$ suppression. 
This makes the harmonic function $S_1$ the only source of the $\log N$ behaviour 
in \eqref{eq:S1only1}, at least at the three-loop level. 

This feature of the evolution kernel \eqref{eq:S1only1} is in a marked contrast with 
the anomalous dimension {\em per se}, whose large $N$ expansion includes growing powers 
of  $\log N$:
\begin{equation}\label{eq:Nexp}
 \gamma(N) = a\,\ln N + \sum_{k=0}^\infty\frac{1}{N^k}\sum_{m=0}^k a_{k,m}\,\ln^m N .
\end{equation} 
Physically, the reduction of singularity of the large $N$ expansion is due to the fact that  
towers of subleading logarithmic singularities in the anomalous dimension are actually 
{\em inherited}\/ from the first loop --- the LBK-classical 
$\gamma^{(1)}=\cP^{(1)} \propto S_1$, 
and the RR evolution equation \eqref{eq:gamma-P} 
generates them automatically\footnote{for the space-like case under consideration, $\sigma=-1$}  \cite{Dokshitzer:2005bf,Dokshitzer:2006nm}.  

\medskip

In this letter we consider twist-3 operators in the $\mathfrak{sl}(2)$ sector 
of ${\cal N}\!=\!4$ SYM and construct the evolution kernel $\cP$ that generates 
the minimal anomalous dimension of single trace operators built of three scalar fields.
This evolution kernel $\cP$ satisfies the Gribov--Lipatov reciprocity,  
with $x^2$ substituted for $x$ in eq.\ \eqref{eq:GLR}, in accord with the fact that 
the harmonic functions entering the twist-3 anomalous dimension have $N/2$ for
the argument \cite{Beccaria:2007,KLRSV:2007}. 
We also demonstrate that the twist-3 kernel $\cP$ admits the representation 
\eqref{eq:S1only1}, with $\hA=\cO{\alpha^4}$. 

This shows that the reciprocity respecting evolution, and thus the property 
of parity preserving series, manifest themselves also beyond the leading twist.

\section{Reciprocity Respecting evolution kernel }

\subsection{Anomalous dimensions of twist-3 operators}

The $\mathfrak{sl}(2)$ sector of planar $\mathcal{N}\!=\!4$ SYM contains single trace states which are linear combinations of the basic operators
\begin{equation} 
\label{eq:sl2states}
\mbox{Tr}\left\{ \left({\cal D}^{s_1}\,Z\right)\ \cdots
\left({\cal D}^{s_L}\,Z\right) \right\},\quad s_1 + \cdots + s_L = N,
\end{equation} 
where $Z$ is one of the three complex scalar fields and ${\cal D}$ is a light-cone covariant derivative. 
The numbers $\{s_i\}$ are non-negative integers and $N$ is the total spin. 
The number $L$ of $Z$ fields is the twist of the operator, {\em i.e.}\/ 
the classical dimension minus spin.
The subsector of states with fixed spin and twist is perturbatively closed under renormalization mixing.

The anomalous dimensions 
of states \eqref{eq:sl2states} are the eigenvalues  $\gamma_L(N; g)$ 
of the dilatation operator --- integrable Hamiltonian. 
These values were obtained by solving numerically the Bethe Ansatz equations (BAE), 
order by order in $g^2$, and guessing the answer in terms of harmonic sums of
transcedentality $\tau=2\,n-1$, with $n$ the number of loops. 
Since wrapping problems, delayed by supersymmetry, appear at $L\!+\!2$ loop order 
for twist-$L$ operators~\cite{Beisert:2005fw,Beisert:2005tm},
the Bethe Ansatz equations for twist-3 
are reliable up to {\em four loops}\/ (including, at the fourth loop, the dressing factor). 

One finds  \cite{Beccaria:2007,KLRSV:2007}
\begin{subequations}\label{eq:answer1234}
\begin{eqnarray}
\label{eq:answer1}
\gamma_{3}^{(1)} &=& 4\, S_1\, ,  \\
\label{eq:answer2}
\gamma_{3}^{(2)} &=& -2\,(S_3+2\,S_1 S_2)\\
\label{eq:answer3}
\gamma_{3}^{(3)} &=& 5\,S_5+6\,S_2\,S_3-8\,S_{3,1,1}+4\,S_{4,1}-4\,S_{2,3}  
 + S_1(4\,S_2^2+2\,S_4+8\,S_{3,1}), { }\qquad  \\
\label{eq:4guess}
\gamma_{3}^{(4)} &=& 
\tfrac{1}{2} \,S_{7}+7 \,S_{1,6}+15 \,S_{2,5}-5 \,S_{3,4}-29 
\,S_{4,3}-21 \,S_{5,2}-5 \,S_{6,1} \nonumber \\
&& -40 \,S_{1,1,5}-32 \,S_{1,2,4}+24 
\,S_{1,3,3}+32 \,S_{1,4,2}-32 \,S_{2,1,4}+20 \,S_{2,2,3} { }\qquad \nonumber \\
&& +40 
\,S_{2,3,2}+4 \,S_{2,4,1}+24 \,S_{3,1,3}+44 \,S_{3,2,2}+24 
\,S_{3,3,1}+36 \,S_{4,1,2}\nonumber \\
&& +36 \,S_{4,2,1}  +24 \,S_{5,1,1}+80 
\,S_{1,1,1,4}-16 \,S_{1,1,3,2}+32 \,S_{1,1,4,1} \\
&& -24 \,S_{1,2,2,2}+16  \,S_{1,2,3,1} -24 \,S_{1,3,1,2}-24 \,S_{1,3,2,1}-24 \,S_{1,4,1,1}
 \nonumber \\
&& -24 \,S_{2,1,2,2} +16 \,S_{2,1,3,1}-24 \,S_{2,2,1,2} -24 \,S_{2,2,2,1}-24 
\,S_{2,3,1,1}\nonumber \\
&& -24 \,S_{3,1,1,2}-24 \,S_{3,1,2,1} -24 \,S_{3,2,1,1}-24 
\,S_{4,1,1,1}-64 \,S_{1,1,1,3,1} \nonumber \\
&&  -8\,\beta
\,S_1\,S_3.  \nonumber 
\end{eqnarray}\end{subequations}
The last term in \eqref{eq:4guess}, with $\beta= \zeta_3$, is the contribution from 
the dressing factor that appears in the BAE at the fourth loop~\cite{Beisert:2006ez}. 

The twist-3 anomalous dimension has two characteristic features: 
\begin{enumerate}
\item 
All harmonic functions $S_{\vec a}$  are evaluated at half the spin,
$ S_\mathbf{a} \equiv S_\mathbf{a}\left({N}/{2}\right)$. 
On the integrability side, this does not look unwarranted, since only 
{\em even}\/ $N$ belong to the non-degenerate ground state of the magnet. 
\item
 No negative indices appear in \eqref{eq:answer1234}, 
 while in the case of twist-2 negative index sums were present starting
 from the second loop \cite{Kotikov:2004er}.
\end{enumerate}
At the $N\to\infty$ limit, the {\em minimal}\/ anomalous dimension $\gamma$ 
(corresponding to the ground state)
must exhibit the universal  (LBK-classical) $\ln N$ behaviour 
which depends neither on the twist, 
nor on the nature of fields under consideration \cite{Braun:1999te}. 
Computing analytically the large $N$ expansion \eqref{eq:Nexp} yields 
(for $\beta= \zeta_3$)
\begin{eqnarray}\label{eq:cusp1}
{\gamma(N)} &\simeq& 4\,\gph^2\> {\ln N}, \nonumber \\
\gph^2 &\equiv& \frac{N_c\,\alp}{2\pi} = g^2 - \zeta_2\, g^4 + \tfrac{11}{5}\zeta_2^2\, g^6 
-  \big(\tfrac{73}{10}\zeta_2^3 +\zeta_3^2\big)\,g^8 + \ldots ,
\end{eqnarray}
which matches the four-loop cusp anomalous dimension 
\cite{Beisert:2006ez,Bern:2006ew}.
This is a non-trivial check, since the derivation of the expressions 
\eqref{eq:answer1234}
was based on experimenting with finite values of the spin $N$.   

Expanding \eqref{eq:gamma-P} gives
\begin{subequations}\begin{equation}\label{eq:gamma-of-P}
\gamma = \cP + \half\cP\dcP +  \tfrac14\bigl[  \cP  \dcP^{2} 
+ \half \cP^{2} \ddcP    \bigr] + 
\tfrac18  \bigl[ \cP \dcP^3 + \tfrac32
 \cP^2\dcP\ddcP + \textstyle{\frac16}\cP^3\dddcP \bigr]
 + \cO{g^{10}} 
\end{equation}
and, equivalently,
\begin{equation}\label{eq:P-of-gamma}
\cP = \gone - \half\gone\dgone +  \tfrac14\bigl[  \gone  \dgone^{2} 
+ \half \gone^{2} \ddgone    \bigr] - 
\tfrac18  \bigl[ \gone \dgone^3 + \tfrac32
 \gone^2\dgone\ddgone + \textstyle{\frac16}\gone^3\dddgone \bigr]
 + \cO{g^{10}} ,
\end{equation}\end{subequations}
where each dot marks derivative over $N$. 

Substituting $\gamma$ and $\cP$ in terms of 
perturbative series in $g^2=\frac{N_c\alpha}{2\pi}$,
\[
 \gamma = \sum_{n=1} g^{2\,n}\, \gamma^{(n)},  
\qquad  \cP = \sum_{n=1}  g^{2\,n}\, {P}^{(n)} , 
\]
after a short calculation one obtains
\begin{subequations} \label{eq:P1234}\begin{eqnarray}
 \label{eq:P1}
P^{(1)} &=& 4\,S_1,  \\
 \label{eq:P2}
P^{(2)}  &=& -2\,S_3-4\,\zeta_2\, S_1, \\
 \label{eq:P3}
P^{(3)}  &=&  S_5 + 2\,\zeta_2\,S_3 + 4\,\left(S_{3,2}+S_{4,1}-2\,S_{3,1,1}\right) \nonumber \\
&&  +\> 4\,S_1\,(2\,S_{3,1}-S_4 + 4\,\zeta_4)
 -4\,S_1^2\,(S_3-\zeta_3) .
\end{eqnarray}
The fourth loop kernel we split into two terms, the first built up of harmonic 
functions $\cS$ only, and the second one containing the zeta-function factors, 
\begin{eqnarray} \label{eq:P4}
 P^{(4)}  &= &   P^{(4)}_S \>+\> P^{(4)}_\zeta . 
\end{eqnarray}
Here we present the corresponding expressions in terms of nested harmonic sums 
with {\em unity}\/ indices moved to the head of the index vector, 
which form turns out to be more compact and is better suited for further 
transformation: 
\begin{eqnarray} \label{eq:P4s}
P^{(4)}_S &=& 8\, \big[  - (S_{3, 3} + S_{1, 5} +2 S_{2, 4}) +4( S_{2, 1, 3}
 + S_{1, 2, 3}+ S_{1, 1, 4}) -8 S_{1, 1, 1, 3}\big]\, S_{1}{ }\qquad   \nonumber\\
&+& \tfrac32\, S_7 -16 \,\big(S_{1,6} + S_{4,3}\big)  -24\, \big(S_{2, 5} + S_{3, 4}\big) 
 \nonumber\\
&& +\, 48\, \big(S_{1,1,5} + S_{1, 3, 3} + S_{3, 1, 3}\big) 
+64\, \big(S_{2,2,3}  + S_{2, 1, 4} +S_{1, 2, 4} \big) \\
&& -\,128\, \big( S_{1, 1, 1, 4} + S_{2, 1, 1, 3}
 + S_{1, 2, 1, 3} + S_{1, 1, 2, 3}\big)  + 256\, S_{1, 1, 1, 1, 3} ;\nonumber \\
 \label{eq:P4z}
 P^{(4)}_\zeta &=& 
8\zeta_4\, \cS_1^3 - 4\big[\zeta_2\zeta_3+8\zeta_5\big]\, \cS_1^2 
-\big[4(\zeta_3+2\beta)\cS_3 + 49\zeta_6\big]\,\cS_1 \nonumber \\
&& +\, (8\cS_{1,1,3}-4\cS_{1,4}-4\cS_{2,3}-\cS_5)\,\zeta_2 -8\cS_3\,\zeta_4 .
\end{eqnarray}\end{subequations}

\subsection{RR harmonic functions}

In order to show that the twist-3 evolution kernel \eqref{eq:P1234}
satisfies the Gribov--Lipatov reciprocity, 
we need to introduce the corresponding basis of harmonic functions. 

The functions 
\[
   \frac{x}{x-1} \ln^{2k}x  \>=\> \Gamma(2k+1)\cdot  \tilde{\cS}_{2k+1}(x)
\]
satisfy the GL reciprocity \eqref{eq:GLR} and generate harmonic functions with
an odd index,
\begin{subequations}\begin{eqnarray}
\frac{1}{\Gamma(2k+1)}\> \Mellin{x \left(\frac{\ln^{2k}x}{x-1}\right)_+}  
&=& \cS_{2k+1}(N) ; \\
\label{eq:hat}
\frac{1}{\Gamma(2k+1)}\> \Mellin{ \frac{x}{x-1} \ln^{2k}x} &=& \hcS{2k+1}(N)
\equiv \cS_{2k+1}(N) - \cS_{2k+1}(\infty) .
\end{eqnarray}\end{subequations}
In \eqref{eq:hat} and hereafter the hat marks the function in the Mellin moment space
with its value at $N\!=\!\infty$ subtracted: $\hat{a}(N)\equiv a(N)-a(\infty)$.

The inverse Mellin transforms $\tcS{\vec{m}}(x)$ of multi-index sums $\cS_{\vec{m}}(N)$ 
that enter eqs.\ \eqref{eq:P3}--\eqref{eq:P4z}  
mix under the reciprocity operation \eqref{eq:GLR}, so  we need to construct 
linear combinations of harmonic functions $\cS$ that do have definite GL parity.

\subsubsection{Integral representation}

An index vector $\vec{m}=\{m_1,m_2,\ldots, m_\ell\}$ corresponds to 
transcedentality $\tau\!=\!\sum_{i=1}^\ell |m_i|$.
We will refer to the transcedentality minus the length of the sum, $w[\vec{m}]=\tau-\ell$,  
as the {\em weight}\/ of the harmonic function.  

Consider the recurrence relation
\begin{subequations}\label{eq:tPhis}\begin{equation}\label{eq:phiint}
 \tPhi{b,\vec{m}}(x) = -[\Gamma(b)]^{-1}
 \frac{x}{x-1} \int_x^1 \frac{dz\,(z+1)}{z^2}
  \,\ln^{b-1}\frac{z}{x} \cdot  \tPhi{\vec{m}}(z) ,
\end{equation}
where the single index function coincides with the image of the standard harmonic sum,
\begin{equation}\label{eq:phi1ind}
 \tPhi{a}(x) =  [\Gamma(a)]^{-1}\,\frac{x}{x-1}\, \ln^{a-1}\frac{1}{x}\>=\> \tcS{a}(x).
\end{equation}\end{subequations}
At the base of the recursion, \eqref{eq:phi1ind}, we have 
\[
  \tPhi{a}(x) \>=\> \bigg(-x\, \tPhi{a}(x^{-1})\bigg) \cdot (-1)^{a-1} \equiv   \bigg(-x\, \tPhi{a}(x^{-1})\bigg) \cdot (-1)^{w[a]}.
\]
An iteration \eqref{eq:phiint} increases the transcedentality of the function by $b$, 
and the length of the index vector by one, so that
\[
 w[\vec{m}]+b-1 = w[b,\vec{m}] .
\]
Observing that the integration measure in \eqref{eq:phiint} transforms 
under $z\to1/z$ as
\[
 d\phi(z) = \frac{dz\,(1+z)}{z^2}, \qquad d\phi(z^{-1}) = -z\, d\phi(z) , 
\]
we conclude, by induction, that the functions $\tPhi{\vec{m}}(x)$ that the equation 
\eqref{eq:tPhis} generates, have definite GL parity determined by their weight:
\begin{equation}\label{eq:Phi-parity-1}
  \tPhi{\vec{m}}(x)\>=\> (-1)^{w[\vec{m}]}  \cdot \bigg( -x\, \tPhi{\vec{m}}(x^{-1}) \bigg).
\end{equation}
Let us represent the r.h.s.\ of \eqref{eq:phiint} as
\begin{equation}  
 \frac1{\Gamma(b)} \frac{x}{x-1} \int_x^1 \frac{dz}{z}
 \,\ln^{b-1}\frac{z}{x} \cdot \frac{z-1}{z}\, \tPhi{\vec{m}}(z) 
 \> - \frac2{\Gamma(b)} \frac{x}{x-1} \int_x^1 \frac{dz}{z}
 \,\ln^{b-1}\frac{z}{x} \, \tPhi{\vec{m}}(z)  .
\end{equation}
Acting on a complementary harmonic sum\footnote{See Appendix for the definition and
properties of complementary harmonic sums} 
$\tucS{a,\vec{n}}$
the first integral produces $\tucS{a+b,\vec{n}}$ while the second generates 
($-2$ times) the sum $\tucS{a,b,\vec{n}}$,   see \eqref{eq:tucSdef}. 
Iterating \eqref{eq:tPhis}, 
\begin{subequations}\begin{eqnarray}
 \tPhi{c,d} &=&  \tucS{c+d} -2\,\tucS{c,d} , \\
 \tPhi{b,c,d} &=&   \tucS{b+c+d} - 2\,( \tucS{b+c,d} +\, \tucS{b,c+d} )+ 4\, \tucS{b,c,d} 
 , \\
 \tPhi{a,b,c,d} &=& \tucS{a+b+c+d}
  - 2\, (\tucS{a+b+c,d}  + \tucS{a+b,c+d} +  \tucS{a,b+c+d} ) \nonumber\\
 &&  +\>4\, (\tucS{a+b,c,d} +\tucS{a,b+c,d} + \tucS{a,b,c+d})
  -8\, \tucS{a,b,c,d}\, , \qquad  \mbox{etc.} 
 \end{eqnarray}\end{subequations}
In the Mellin moment space,  the formal construction of functions $\Phi_{\vec{m}}(N)$ 
in terms of complementary harmonic sums is described in Appendix, see \eqref{eq:ydef},
\eqref{eq:Phidef}.

Equation \eqref{eq:Phi-parity-1} shows that
the functions $\tPhi{\vec{m}}$ with {\em even weight}\/ $w[\vec{m}]=\tau-\ell$ are 
Reciprocity Respecting (RR). 

\subsection{Answer}

In terms of the functions $\Phi$ so introduced 
the evolution kernel \eqref{eq:P1234} reads 
\begin{subequations} \label{eq:Pnonph}
\begin{eqnarray}
 P^{(1)} &=& 4\,\cS_1, \\
 P^{(2)} &=& - 4\,\zeta_2\,\cS_1 -2\cS_3, \\
 P^{(3)} &=& \tfrac{44}5\, \zeta_2^2\, \cS_1 + 2\zeta_2 \,\cS_3 
 + 3\,\cS_5 -2\,\Phi_{1,1,3},\\
 P^{(4)} &=& 
 \big(\tfrac{146}{5}\zeta_2^3 +4\zeta_3^2\big)\, \cS_1   
 -\zeta_2\,  \big( 3\,\cS_5 -2\,\Phi_{1,1,3}\big)
 -\tfrac{24}{5}\, \zeta_2^2\, \cS_{3} + 4\,S_1\, \widehat{\cal {A}}_4 + {\cal B}_4  , 
  { }\qquad 
\end{eqnarray}
\end{subequations}
where
\begin{subequations}\label{eq:AB-4}
\begin{eqnarray}\label{eq:calA4}
\widehat{\cal A}_4 &=& 2\,\hPhi{1,1,1,3} -\,(\hPhi{1,5}+\hPhi{3,3}) 
-(2\beta-\zeta_3)\, \hcS{3}, \\
\label{eq:calB4}
{\cal B}_4 &=& 16\,\Phi_{1,1,1,1,3} -4\big( \Phi_{3,1,3}+  \Phi_{1,3,3}
 +  \Phi_{1,1,5} \big)  - \tfrac{5}{2}\,\cS_7 ,
\end{eqnarray}\end{subequations}
We recall that the arguments of the functions on the r.h.s.\ of the equations is $N/2$, 
and $\beta=\zeta_3$ in \eqref{eq:calA4}.
Alternatively, in terms of the physical coupling \eqref{eq:cusp1},
the perturbative series for the kernel, 
$   \cP = \sum_{n=1}\gph^{2n} \> {\cP}_{\ph}^{(n)} $, 
takes the compact form 
\begin{subequations}\label{eq:ans-1234}\begin{eqnarray}\label{eq:ans-1}
 \cP^{(1)}_\ph &=& 4\,\cS_1, \\
\label{eq:ans-2} \cP^{(2)}_\ph &=&  -2\,\cS_3, \\
\label{eq:ans-3} \cP^{(3)}_\ph &=& 3\,\cS_5 -2\,\Phi_{1,1,3}\>
+\> \zeta_2 \cdot (-2\,\cS_3)  ,\\
\label{eq:ans-4}
 \cP^{(4)}_\ph &=&   4\,S_1\cdot  \widehat{\cal {A}}_4 +  {\cal B}_4 
\>+\>  2\,\zeta_2 \cdot \big( 3\,\cS_5 -2\,\Phi_{1,1,3}\big) .{ }\qquad 
\end{eqnarray}\end{subequations}
Since all harmonic functions involved have {\em even}\/ weights $w$, 
the splitting kernel \eqref{eq:ans-1234} is Reciprocity Respecting.
Mark that given the argument $N/2$ of the harmonic functions in \eqref{eq:ans-1234},
 \eqref{eq:AB-4}, the reciprocity relation \eqref{eq:GLR} applies to $\tcP(x^2)$.

This result can be compared with the evolution 
kernel\footnote{the GL parity of a harmonic function with $k$ negative 
indices is $(-1)^{w+k}$ } that generates 
the twist-2 universal anomalous dimension~\cite{Dokshitzer:2006nm}: 
\begin{subequations}\label{eq:ans-t2}\begin{eqnarray}
 \cP^{(1)}_\ph &=& 4\,\cS_1(N); \\
 \cP^{(2)}_\ph &=& -4\, \cS_{3}(N) +4\, \Phi_{1,-2}(N)  ;\\
 \cP^{(3)}_\ph &=&  8\, \cS_{5}(N) -  24\, \Phi_{1,1,1,-2}(N)  -8\,\zeta_2\,  \cS_{3}(N) \nonumber\\
 &-& 8\,\cS_1(N)\cdot \big[ 2\,\hPhi{1,1,-2}(N) + \hPhi{-2,-2}(N)
 - \hcS{-4}(N) + \zeta_2\,  \hcS{-2}(N)  \big] .
\end{eqnarray}\end{subequations}

\subsection{Asymptotic expansion}

Large Mellin moments $N$ correspond to parton light-cone momentum 
fractions $x\to 1$. In this region only partons with small energy--momentum can be 
produced in the final state of a hard process, $(1-x)\ll1$, and radiation of soft gluons 
dominates the answer.
Therefore, the large $N$ behaviour is important for understanding 
how the LBK physics manifests itself in the structure of anomalous dimensions and 
of the evolution kernel. 

\medskip

At large positive values of the argument, $z\to +\infty$,  
the harmonic function $\cS_1(z)$ has the asymptotic expansion
\begin{subequations}\label{eq:asym}\begin{equation}\label{eq:S1asym}
S_1(z)= \psi(z+1)-\psi(1) = (\ln y +\gamma_E)  + \tfrac16 y^{-2} - \tfrac1{30} y^{-4}
+ \tfrac 4{315} y^{-6} + \ldots ,
\end{equation}
where
\begin{equation}
 y^2 = z(z+1) . 
\end{equation}\end{subequations}
Such a structure of the series is inherited by all harmonic sums $\cS_a(z)$, 
which can be obtained from \eqref{eq:asym} by simple differentiation,
\[
 \hcS{a}(z) = \frac{(-1)^{a-1}}{\Gamma(a)} \left(\frac{d}{dz}\right)^{a-1} \cS_1(z), \qquad  
  \frac{d}{dz} = \frac{dy}{dz}\,\frac{d}{dy} =  \sqrt{1+\frac1{4\,y^2}}\,\frac{d}{dy} ,
\]
as well as by the nested harmonic functions $\Phi_{\vec{m}}(z)$.
Namely, the asymptotic series of the functions
of even (odd) weight contain only even (odd) inverse powers of $y$.  
This is a direct consequence of their inverse Mellin images having definite GL-parity 
\cite{Basso:2006nk}.

\subsubsection{Conformal structure of the large $N$ expansion}

This feature of the asymptotic expansion has a clear symmetry origin \cite{Basso:2006nk}.

On the light-cone the residual conformal invariance is that of the collinear 
subgroup of the conformal group $SL(2;\mathbb{R})\subset SO(2,4)$~\cite{O81}.
In the $\mathcal{N}=4$ SYM theory, (super)conformal invariance is exact at all loops
and quasipartonic operators can be classified according to the representations of $SL(2;\mathbb{R})$.
Also, the operators belonging to different $SL(2;\mathbb{R})$ multiplets 
do not mix under renormalization. The 
anomalous dimension of each conformal multiplet depends only on the conformal spin characterizing the representation
\begin{equation}
j(g) = \tfrac{1}{2}(N + \Delta(g)) = N + \tfrac12\,{L} + \tfrac{1}{2}\,\gamma_L(N),
\end{equation}
where $\Delta(g)$ is the scaling dimension, $\Delta= N+L+\gamma_L$, 
which includes the quantum anomalous contribution $\gamma_L(N;g)$. 
Hence, the spin dependence of the anomalous dimension at fixed twist 
$L$ is encoded in the non-linear relation
\begin{equation}
\label{eq:non-linear}
\gamma_L(N) = f_L\left(N + \tfrac{1}{2}\,\gamma_L(N)\right),
\end{equation}
which coincides with the relation \eqref{eq:gamma-P} 
between the anomalous dimension and the 
Reciprocity Respecting evolution kernel $\cP$ for the space-like evolution, $\sigma=-1$.

As we have mentioned in the Introduction, the fact that the asymptotic series for the 
RR evolution kernel runs in integer powers of $y^2$ (modulo logarithmic corrections)
has been dubbed ``parity preserving asymptotic series'' and verified 
for a broad variety of QFT models in \cite{Basso:2006nk} 
for the twist-2 case.
Now we see that this property holds for twist-3 as well.  

In general, the conformal Casimir operator reads~\cite{K96,Belitsky:2006en}
 \begin{eqnarray}\label{eq:J2}
  J^2 = (N+L\,\eta-1)(N+L\,\eta),
 \end{eqnarray}
where $\eta = \frac12,1,\tfrac32$ for scalars, gaugino, and gauge boson fields, respectively.

For twist-2, the argument of harmonic functions in eqs.\ \eqref{eq:ans-t2} coincides with the Lorentz spin, $z=N$,  and we have simply $J^2= N(N+1)=y^2$.
For the twist-3 in the scalar sector, \eqref{eq:ans-1234}, 
we have $z=N/2$; substituting $L=3$, $\eta=\tfrac12$ into \eqref{eq:J2} gives
\[
J^2= (N+\tfrac12)(N+\tfrac32) = 4\, (\tfrac{N}{2}+\tfrac14) (\tfrac{N}{2}+\tfrac34) 
=  4\,z^2 + 4\,z + \tfrac34 = 4\,y^2+\tfrac34.
\] 
Thus, the expansion in $y^2$ translates into that in the quadratic Casimir, $J^2$. 

Asymptotic series for the evolution kernel, both for the twist-2 and twist-3 operators,
can be cast in the form
\begin{equation}   \label{eq:Pser}
\cP(N) = (\ln y\!+\!\gamma_E) \left[4\,\gph^2 
+\! \sum_{n=1}^\infty \frac{a_n}{y^{\,2\,n}} \right]
 +\!  \sum_{m=0}^\infty  \frac{ b_{m} }{y^{\,2\,m} }\,, 
  \quad  \left\{  \begin{array}{ll}  y^2= N(N+1), & L\!=\!2; \\ 
  y^2 = \tfrac{N}{2}(\tfrac{N}{2}+1) , & L\!=\!3. 
  \end{array}\right. { }\quad
\end{equation}
where the $L$-dependent coefficients $a$, $b$ are given in series of the coupling.

\subsubsection{Logarithmic structure of the large $N$ expansion}

Apart from ``parity preservation'', the expansion \eqref{eq:Pser} has another remarkable 
feature. In a sharp contrast with the series 
for the {\em anomalous dimension}, where the number of logs increases with the power
of the $1/N$ suppression, see  \eqref{eq:Nexp}, the {\em evolution kernel}\/ appears 
to be {\em linear}\/ in $\ln N$. 

Whether this property holds beyond the known orders ($g^6$ for $L=2$; 
$g^8$ for $L=3$) remains unknown. 
Should this be the case, that would mean that the tower of subleading logarithmic singularities in the anomalous dimension are actually 
{\em inherited}, in all orders, from the first loop --- the LBK-classical 
$\gamma^{(1)}=\cP^{(1)} \propto S_1$, and the RR evolution equation 
\eqref{eq:gamma-P} generates them automatically.  

The leading logarithmic behaviour is universal and is given, in terms of the physical coupling \eqref{eq:cusp1},  by the  LBK-classical contribution to the evolution kernel,
\begin{equation}\label{eq:Plead}
 \cP(N) \>\simeq\> 4\,\gph^2\, \cS_{1}(z) = 4\,\gph^2 \ln N \>+\> \cO{1}.
\end{equation}
The non-trivial $N^2$ dependence accompanying the $\log N$ enhancement, 
$\delta\cP \!\propto\! \log N/N^2$, appears only at the level of $a_1\propto g^6$ and
$a_1\propto g^8$ for twist-2 and 3, correspondingly.

\medskip

It follows from the structure of the relation
\eqref{eq:gamma-of-P},
\begin{equation}\label{eq:BKexp}
  \gamma(N) \>=\> \sum_{k=1}^\infty \frac1{k!}\left( \frac12\,\frac{d}{dN}\right)^{k-1} 
  \big[ \cP(N)\big]^k,
\end{equation}
that the maximally logarithmically enhanced contributions to the anomalous dimension,
 $\log^{k} N/N^k$,  
are driven by the leading, 
first loop term \eqref{eq:Plead}. 
Representing the twist-3 anomalous dimension as 
\begin{equation}\label{eq:gamma3exp}
 \gamma(N; L\!=\!3) 
 = a_0\,\big(\ln \frac{N}{2}+\gamma_E\big) + B 
 + \sum_{k=1}^\infty\frac{1}{N^k}\sum_{m=0}^k a_{k,m}\,\ln^m N ,\qquad a_0 = 4\,\gph^2,
\end{equation} 
and substituting \eqref{eq:Plead} into \eqref{eq:BKexp} we immediately obtain
\begin{equation}\label{eq:akk}
 a_{1,1} = \tfrac12 a_0^2\,,\quad  a_{2,2} = -\tfrac18a_0^3\, ,\quad  
 a_{3,3} = \tfrac1{24}a_0^4\,; \qquad
 a_{k,k} =  \frac{(-1)^{k-1}}{2^k\,k}\, a_0^{k+1}.
\end{equation}
In order to predict the next-to-maximal logarithmic terms, $a_{k,k-1}\log^{k-1} N/N^k$, 
in the expansion \eqref{eq:gamma3exp},
it suffices to keep the constant in \eqref{eq:Plead} and use the approximation   
\begin{subequations}\begin{equation}\label{eq:Psublead}
 \begin{split}
 \cP(N) \>&\simeq 
 4\,\gph^2\, \cS_{1}(z) + B = 4\,\gph^2\cdot \left(\ln \sqrt{z(z+1)} +\gamma_E\right) + B 
 + \cO{z^{-2}}\\ 
 & \simeq 4\,\gph^2 \left[\ln z +\gamma_E + \frac{1}{2z}\right] + B,
\end{split}
\end{equation}
where
\begin{equation}
 B = B(g^2,L\!=\!3) = -2\zeta_3\,\gph^4  -(2\zeta_2\zeta_3+\zeta_5)\,\gph^6  
 +  \cO{\gph^8} . 
\end{equation} \end{subequations}
In particular, this gives (recall that for twist-3 we have $z=N/2$)
\begin{equation}\label{eq:akkmo}
 a_{1,0}\>=\> a_0\big(1 +  \tfrac12 B\big) .
\end{equation}
Relations \eqref{eq:akk},  \eqref{eq:akkmo} hold in all orders in the coupling constant.

\section{Conclusions \label{Sec-Conc}}

The notion of the {\em reciprocity respecting}\/ evolution equation (RREE) emerged 
in an attempt to combine, in a single framework, anomalous dimensions 
of space-like parton distributions and time-like parton fragmentation functions,
in order to simplify the structure of the higher order corrections.  
The basic observation is that the complexity of higher loop contributions 
is, to a large extent, {\em inherited}\/ from lower orders. 
This is especially so for the major part of the QCD anomalous dimensions 
which is governed by the ``classical'' gluon radiation,  
in the sense of the Low--Burnett--Kroll (LBK) theorem. 
One may argue that there should exist a framework in which effects of the classical 
gluon fields would be fully generated, in all orders, from the {\em first loop}, 
in the spirit of the LBK wisdom. 

In the space of Mellin moments of parton distributions, 
the corresponding part of the anomalous dimension, 
in the $n$th order of the perturbative loop expansion, 
is described by harmonic functions of the {\em maximal transcedentality}, 
$\tau=2\,n-1$.
 
 \medskip

Since QCD shares the gauge boson sector with supersymmetric theories, 
the latter provide a perfect ground for testing the inheritance idea. 
The maximally supersymmetric ${\cal{N}}\!=\!4$ SYM model is exceptional in this respect.
In this theory {\em quantum effects}\/ due to gluons, gauginos and scalars cancel 
in the beta-function (in all orders) as well as in the one-loop anomalous dimension. 
 
\medskip

In this paper we have addressed the question of applicability of the notion of RREE 
beyond the leading twist.
We considered the bosonic $sl(2)$ sector of the ${\cal{N}}\!=\!4$ SYM and 
found that the Gribov--Lipatov reciprocity relation holds in four loops 
for the evolution kernel $\cP$ describing the minimal (ground state) 
anomalous dimension of twist-3 operators built of three scalar fields. 
The GL relation holds with $x^2$ substituted for $x$, in accord with
the fact that the harmonic functions entering the twist-3 anomalous
dimension have $N/2$ as the argument.

\medskip

The twist-3 evolution kernel is given in eqs.\  \eqref{eq:ans-1234}, \eqref{eq:AB-4}. 
Having derived this kernel from the space-like anomalous dimension, 
prediction for the {\em time-like}\/ twist-3 anomalous dimension follows 
immediately from the RREE. 
It is given by \eqref{eq:gamma-of-P} with the r.h.s.\ changed to 
$-\mbox{r.h.s.}[-\cP]$ (i.e.,
changing signs of the terms even  in $\cP$).

The structure of the twist-3 evolution kernel turns out to be even simpler 
than that for the leading twist-2, \eqref{eq:ans-t2}. 
In particular, an extreme simplicity of the second loop kernel \eqref{eq:ans-2}
is likely to be due to the fact that the genuine $\cO{\alpha^2}$ graphs
that renormalise twist-3 operators {\em vanish}\/ identically because of the colour  
structure (each pair in the adjoint representation): 
\[
 \epsfig{file=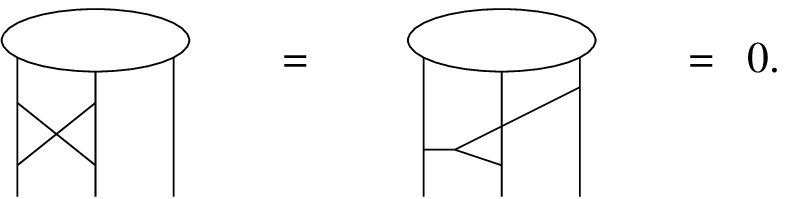}
\]

In the large $N$ limit, where classical LBK gluon radiation dominates, 
the asymptotic series for evolution kernels are much less singular than the corresponding
series for the anomalous dimensions. 
Namely, only the first power of $S_1\propto \ln N$ is present in the evolution kernel, 
both for twist-3 \eqref{eq:ans-1234}, and twist-2 \eqref{eq:ans-t2}. 
Whether this remarkable property is of general nature and holds in higher orders 
($n\ge 4$ for twist-2, and $n\ge 5$ for twist-3) remains unknown.

In twist-3, the subleading logarithmic enhancement 
$\cS_1\cdot \widehat{\cal A}_4(N)\propto \ln N/N^2$ appears for the first time
at the level of $\alpha^4$, while in the case of twist-2 such term is present in the third loop, $\sim \alpha^3\,\ln N/N^2$.  

\medskip
The specific features of the minimal twist-3 anomalous dimension, 
and of the corresponding evolution kernel, such as the absence of negative indices
and appearance of $N/2$ as the argument of harmonic structures 
are awaiting physical explanation.

\section*{ Acknowledgements }

One of us (YLD) is indebted to Gregory Korchemsky and Benjamin Basso for 
an illuminating discussion.

\appendix	\setcounter{equation}{0}

\section{RR harmonic structures}
\subsection{Complementary harmonic sums}

Let $\vec{m}$ denote a string of integers, 
$\vec{m}=\{m_1,m_2,\ldots , m_\ell \}$.
We define multi-index (or ``nested'') complementary harmonic sums $\ucS{m}$
by the recursive relation
\begin{subequations}\begin{equation}\label{eq:complement-def}
 \ucS{a,\vec{m}}(N) = \cS_{a,\vec{m}}(N) - \cS_a(N)\cdot  \ucS{\vec{m}}(\infty) ,
\end{equation}
where we set
\begin{equation}
 \ucS{a}(N) \equiv \cS_{a}(N),
\end{equation}\end{subequations}
For any index vector $\vec{m}$ with the last index $m_\ell \neq 1$, 
the value $\ucS{\vec{m}}(\infty)$ is finite, 
so that we can carry out subtraction to introduce
\begin{equation}\label{eq:hucS-def}
 \hucS{\vec{m}}(N) \equiv   \ucS{\vec{m}}(N)  -  \ucS{\vec{m}}(\infty) .
\end{equation} 
For positive integer $N$ we have 
\begin{equation}\label{eq:hucS-recurs}
  \hucS{a,\vec{m}}(N) = - \sum_{n=N+1}^\infty (\mathop{{\rm sgn}} a)^n\, 
  n^{-|a|}\, \hucS{\vec{m}}(n). 
\end{equation}
From now on we restrict ourselves to positive 
indices\footnote{Generalisation is straightforward; for complementary sums with 
a negative index see \cite{Dokshitzer:2006nm}.} 
so that for the index vector $\vec{m}\!=\!\{m_1,m_2,\ldots , m_\ell \}$, 
with $m_\ell \ge2$,
\begin{subequations}\label{eq:ucS-sums}
\begin{eqnarray}\label{eq:ucS-sum}
   \ucS{\vec{m}}(N) &=& (-1)^{\ell-1} \sum_{n_1=1}^N n_1^{-m_1}
    \sum_{n_2=n_1+1}^\infty n_2^{-m_2}\>\> \ldots  \!\!
    \sum_{n_\ell=n_{\ell-1}+1}^\infty n_\ell^{-m_\ell} , \\
\label{eq:hucS-sum}
   \hucS{\vec{m}}(N) &=& (-1)^\ell \sum_{n_1=N+1}^\infty n_1^{-m_1}
    \sum_{n_2=n_1+1}^\infty n_2^{-m_2}\>\> \ldots  \!\!
    \sum_{n_\ell=n_{\ell-1}+1}^\infty n_\ell^{-m_\ell} .
\end{eqnarray}\end{subequations}

\subsubsection{Large $N$ behaviour}
As follows from their definition \eqref{eq:ucS-sums}, (subtracted) complementary sums
fall in the $N\to\infty$ limit. 
The leading power of the large $N$ behaviour is given by the {\em weight}\/ of the sum, 
$w$ (equal transcedentality minus length):
\begin{equation}
\hucS{\vec{m}}(N) \propto N^{-w[\vec{m}]}, \qquad w[\vec{m}]\equiv \tau-\ell \equiv \sum_{i=1}^\ell (m_i-1).
\end{equation}
Importantly, large $N$ asymptotic expansion of complementary sums 
contains no $\log N$ enhanced terms, at any order of the $1/N$ suppression.

\subsubsection{Mellin image}
Inverse Mellin images of  complementary sums, 
\begin{equation}
\tucS{\vec{m}}(x)=  \invMellin{\, \hucS{\vec{m}}(N)  \,} ,
\end{equation}
can be generated by the integral operation
\begin{equation}\label{eq:tucSdef}
\tucS{a,\vec{m}}(x) = [\Gamma(a)]^{-1} \frac{x}{x-1}
 \int_x^1 \frac{dz}{z}\, \ln^{a-1} \frac{z}x\cdot \tucS{\vec{m}}(z) . 
\end{equation}
In particular,
\begin{subequations}\begin{eqnarray}
\tucS{b,a}(x) &=& [\Gamma(a)\Gamma(b)]^{-1} \frac{x}{x-1}
 \int_x^1 \frac{dz}{z-1}\, \ln^{b-1} \frac{z}x\> \ln^{a-1}\frac1z , \\
\tucS{c,b,a}(x) &=& \!\! 
[\Gamma(a)\Gamma(b)\Gamma(c)]^{-1} \frac{x}{x\!-\!1}
 \int_x^1 \frac{dy}{y\!-\!1}\, \ln^{c-1} \frac{y}x 
 \int_y^1\frac{dz}{z\!-\!1} \ln^{b-1}\frac{z}{y} \ln^{a-1}\frac1z , {   }\qquad \qquad
\end{eqnarray}\end{subequations}
etc.

\subsubsection{Derivative over $N$}

\begin{equation}
\frac{d}{dN}\, \ucS{\vec{m}}(N)  =  \Mellin{\,\ln x\cdot \tucS{\vec{m}}(x) \,}
 = -\sum_{k=1}^\ell m_k\cdot \hucS{m_1,\ldots, m_k, m_k+1, m_{k+1},\ldots, m_\ell}(N),
\end{equation}
Using the definition \eqref{eq:complement-def}, it is straightforward to derive the
formula for the derivative of the standard nested harmonic sums,
\begin{equation}
\frac{d}{dN}\, \cS_{\vec{m}}(N) 
 = -\sum_{k=1}^\ell m_k\, \hcS{m_1,\ldots, m_k+1,\ldots, m_\ell}(N)
 + m_\ell\sum_{k=1}^\ell \hcS{m_1,\ldots,m_k}(N)\cdot \ucS{m_{k+1},\ldots,m_\ell}(\infty).
\end{equation}

\subsection{Construction}
Given a complementary sum $\ucS{\vec{m}}$ of length $\ell$, we construct 
a combination of sums $y^{(r)}$ 
having the same transcedentality and {\em reduced length}. 
The index vector $\vec{m}$ has $\ell$ indices, separated by $\ell-1$ commas;  
by erasing $(\ell-r)$ of them, 
and summing up the non-separated indices, we construct  
\begin{subequations}\label{eq:ydef}\begin{eqnarray}
 y^{(\ell)}_{\vec{m}} &=& \ucS{\vec{m}} \equiv \ucS{m_1,m_2,m_3,\ldots, m_\ell} , \\
 y^{(\ell-1)}_{\vec{m}}  &=& \ucS{[m_1+m_2],m_3,\ldots, m_\ell} 
 + \ucS{m_1,[m_2+m_3],\ldots, m_\ell} +\ldots + \ucS{m_1,m_2,\ldots, [m_{\ell-1}+m_\ell]} , { }\qquad \\
 y^{(\ell-2)}_{\vec{m}}  &=& \ucS{[m_1+m_2+m_3],m_4,\ldots, m_\ell} 
 + \ucS{[m_1+m_2],[m_3+m_4],\ldots, m_\ell} \nonumber \\ 
 && + \ucS{[m_1+m_2],m_3,[m_4+m_5],\ldots, m_\ell} 
 +\ldots + \ucS{m_1,m_2,\ldots, [m_{\ell-2}+m_{\ell-1}+m_\ell]} ,\\
 &\ldots & \nonumber \\
y^{(2)}_{\vec{m}}  &=& \sum_{k=1}^{\ell-1}\ucS{[\sum_1^k m_i], [\sum_{k+1}^\ell m_j]} , \\  
y^{(1)}_{\vec{m}}  &=& \ucS{[\sum_1^\ell m_i]} =  \cS_{\tau}.  
\end{eqnarray}\end{subequations}
Now, we consider definite combinations of complementary harmonic sums:
\begin{equation}\label{eq:Phidef}
 \Phi_{\vec{m}} = \sum_{r=1}^{\ell} (-2)^{r-1} y^{(r)}_{\vec{m}}
 =(-2)^{\ell-1} \ucS{\vec{m}} 
+ \ldots  +  4  y^{(3)}_{\vec{m}} -2 y^{(2)}_{\vec{m}} + \cS_{\tau} . 
\end{equation}
Specificity of the combinations so organised lies in the structure of their
inverse Mellin images. Namely,
inverse Mellin images of (subtracted) harmonic functions \eqref{eq:Phidef},
\begin{equation}
  \tPhi{\vec{m}}(x)= \invMellin{ \hPhi{\vec{m}}(N) } ,
\end{equation}
have definite GL parity, 
\begin{equation}\label{eq:Phi-parity}
  \tPhi{\vec{m}}(x)\>=\> (-1)^w  \cdot \bigg( -x\, \tPhi{\vec{m}}(x^{-1}) \bigg).
\end{equation}
This shows that the functions $\tPhi{\vec{m}}$ 
with {\em even weight}\/ $w[\vec{m}]=\tau-\ell$ are RR. 

\subsubsection{Some useful examples}

{$\tau={\bf 5}$}:
\begin{equation}
\Phi_{1,1,3} = 4\,  \ucS{1,1,3} -2 (\ucS{1,4}+\ucS{2,3}) + \cS_5. 
\end{equation}
{$\tau={\bf 6}$}:
\begin{subequations}\begin{eqnarray}
\Phi_{3,3} &=&  -2\, \ucS{3,3} + \cS_6; \\
\Phi_{1,5} &=&  -2\, \ucS{1,5} + \cS_6; \\
\Phi_{1,1,1,3} &=& - 8\,  \ucS{1,1,1,3} + 4( \ucS{1,1,4} + \ucS{1,2,3} +  \ucS{2,1,3}) 
 -2 (\ucS{1,5}+\ucS{2,4}+\ucS{3,3})  +  \cS_6. {  }\qquad \qquad 
\end{eqnarray}\end{subequations}
{$\tau={\bf 7}$}:
\begin{subequations}\begin{eqnarray}
\Phi_{1,3,3} &=& 4\, \ucS{1,3,3} -2 (\ucS{1,6}+\ucS{4,3}) + \cS_7 ; \\
\Phi_{3,1,3} &=&  4\,  \ucS{3,1,3}  -2 (\ucS{4,3}+\ucS{3,4}) +\cS_7; \\
\Phi_{1,1,5} &=&4\,\ucS{1,1,5}-2 (\ucS{1,6}+\ucS{2,5}) +  \cS_7  ; \\
\Phi_{1,1,1,1,3} &=& 16\,  \ucS{1,1,1,1,3}  - 8 (\ucS{1,1,1,4} +\ucS{1,1,2,3} +\ucS{1,2,1,3} +\ucS{2,1,1,3}) \nonumber \\
 && +\> 4( \ucS{1,1,1,4} + \ucS{1,1,2,3} +  \ucS{1,2,1,3} +\ucS{2,1,1,3}  ) \\
 && -\>2 (\ucS{1,6}+\ucS{2,5}+\ucS{3,4}+\ucS{4,3}) + \cS_7 .
\nonumber{ }\qquad 
\end{eqnarray}\end{subequations}

\end{document}